\documentclass[%
 reprint,
 amsmath,amssymb,
 aps,
]{revtex4-1}

\usepackage{graphicx,setspace}
\usepackage{dcolumn}
\usepackage{bm}
\usepackage[mathlines]{lineno}
\usepackage[colorlinks=true,citecolor=blue,linkcolor=blue,anchorcolor=green, anchorcolor=green,urlcolor=blue]{hyperref}
\usepackage{mathrsfs}
\usepackage{xcolor}
\usepackage{float}
\usepackage{ulem}
\usepackage{lipsum} 

\let\originalleft\left
\let\originalright\right
\renewcommand{\left}{\mathopen{}\mathclose\bgroup\originalleft}
\renewcommand{\right}{\aftergroup\egroup\originalright}
\mathcode`\*="8000
{\catcode`\*=\active\gdef*{\mathclose{}\,\mathopen{}}}

\newcommand{\cu}[1]{\left\{#1\right\}}

\newcommand{\be}{\begin{equation}}
\newcommand{\ee}{\end{equation}}
\newcommand{\bea}{\setlength\arraycolsep{2pt} \begin{eqnarray}}
\newcommand{\eea}{\end{eqnarray}}
\newcommand{\nn}{\nonumber}

\begin{document}
\preprint{APS/123-QED}
\title{Observational signature of near-extremal Kerr-like black hole in a modified gravity theory at the Event Horizon Telescope}

\author{Minyong Guo}
\affiliation{\small
Department of Physics, Beijing Normal University,
Beijing, 100875, China}
\affiliation{\small
Perimeter Institute for Theoretical Physics, Waterloo, Ontario N2L 2Y5, Canada}

\author{Niels A. Obers}
\author{Haopeng Yan}
\affiliation{\small
The Niels Bohr Institute, University of Copenhagen, Blegdamsvej 17, 2100 Copenhagen {\O}, Denmark}

\begin{abstract}
  We study the shadows cast by near-extremal Kerr-MOG black holes for different values of the parameter in modified gravity (MOG). In particular, we consider an isotropic emitter orbiting near such black holes and analytically compute the positions, fluxes and redshift factors of their images. The size of the shadow decreases when the modified parameter is increased. For each shadow, the images of the emitter appear on a special part of the shadow which has a rich structure. The primary image and secondary images are similar to those produced for the near-extremal (high spin) Kerr black hole, but the near-extremal Kerr-MOG black hole can have a spin ($
  \hat{J}/M^2_{\alpha}$) which is finitely lower than 1. When the modified parameter is varied,
  the typical positions of the corresponding images do not change, nor does the typical redshift factor associated with the primary image. However, another typical redshift factor associated with the secondary image increases when the modified parameter is increased. We also find that the fluxes increase in that case. These images appear periodically with period greater than that of Kerr. This provides an alternative signature away from the Kerr case which may be tested by the Event Horizon Telescope.
\end{abstract}

\maketitle




\section{Introduction}

Black holes play an important role both in understanding gravity theories and in explaining astronomical phenomena. There have been abundant observational evidences for black holes in our universe \cite{eckart1996observations, casares2006observational,narayan2013observational}. Inspiringly, we are entering a new era of more precise astronomical observations with the efforts of LIGO, Virgo, the Event Horizon Telescope (EHT), ATHENA, SKA, eLISA, et al \cite{ligo,virgo,eht,athena,ska,elisa}.
Among those, EHT is particularly interesting \linebreak since it aims at observing the event horizon of a black hole which is its most striking feature. Thus we will be able to observe black holes significantly closer to the event horizon and obtain their images (shadows). Hence there is an urgent need for theoretical templates to identify the images that one expects to observe. This has stimulated recent theoretical works  predicting the signals that EHT may possibly observe \cite{lu2014imaging,psaltis2015general,johannsen2016testing,ricarte2014event,
psaltis2016quantitative,gralla2017observational,Lupsasca:2017exc}
and examining the type  of properties of gravity that the shadows can inform us  of \cite{Cunha:2018acu,Cunha:2018gql,Mizuno:2018lxz}.


Recently, an analytical method was proposed to compute the observational signature of a near-extremal (high spin) Kerr black hole \cite{gralla2017observational}. The authors considered an isotropically emitting point source (``hot spot") orbiting near a rapidly spinning Kerr black hole and found that the primary image and secondary images appear on a vertical line segment which constitutes a portion of the black hole shadow. Ref.~\cite{gralla2017observational} also discussed the positions, fluxes and redshift factors of these images in detail, which provide a unique signature for identifying a high spin Kerr black hole if detected by observations.


Even though the Kerr solution predicted by general relativity (GR) is expected to describe astrophysical black holes, there are indications both from physics and astrophysics that GR is modified. Therefore, it is important to obtain templates based on different gravitational theories \cite{sen1992rotating,moura2006higher,ayzenberg2014slowly,
capozziello2008extended,myers1986black,amarilla2010null,abdujabbarov2013shadow,
moffat2006scalar}. One of these candidates is the scalar-tensor-vector (STVG) modified gravitational (MOG) theory \cite{moffat2006scalar}.
The motivation of this theory is to construct a theory of gravity without invoking dark matter, since the hypothesized dark matter has not been observed yet, to release the discrepancies between some observed data and theoretical predictions of GR.
For the same motivation, a series of theories has been proposed prior to the MOG theory, namely, the Modified Newtonian Dynamics theory (MOND) \cite{milgrom1983modification} and its relativistic extensions (for a review see Ref.~\cite{Famaey:2011kh}).
Independently, Moffat proposed the MOG theory by adding a massive vector field to the Einstein-Hilbert action and replacing constants of the ordinary theory by scalar fields \cite{moffat2006scalar}.
The MOG theory has successfully explained Solar system observations \cite{moffat2006scalar}, galactic rotation curves \cite{moffat2013mog}, dynamics of clusters of galaxies \cite{Moffat:2013uaa} and cosmological data \cite{Moffat:2007ju}. However, it still remains to be tested in the strong gravity regime \cite{Perez:2017spz}. The EHT might hopefully provide such a test.

The static and rotating solutions of MOG were obtained in Ref.~\cite{Moffat:2014aja} and followed by research works examining various aspects of these black holes \cite{Mureika:2015sda,Moffat:2015kva,Moffat:2016gkd,Armengol:2016xhu,lee2017innermost,
Perez:2017spz,Sharif:2017owq,Roshan:2018lhu,Manfredi:2018meg}. The particular case of rotating black holes known as Kerr-MOG black holes have gained more astrophysical interests since the observed black holes are thought to be rotating. For example, the particle dynamics \cite{Sharif:2017owq}, the innermost stable circle orbit \cite{lee2017innermost}, the accretion disks \cite{Perez:2017spz} and the relativistic jets \cite{Armengol:2016xhu} have been studied. Furthermore, the shadows cast by MOG black holes have  been studied in Ref.~\cite{Moffat:2015kva}, in which it was shown that the sizes of these shadows increase significantly as the free modified parameter is increased. However, the shadow is expected to exhibit further signals which need to be clarified, such as its shape and images of orbiting hot spots (if any) on it.



The aim of this paper is to obtain the shadow cast by a near-extremal
Kerr-MOG black hole, following the method of Ref.~\cite{gralla2017observational}. Moreover, we will  study the images of an isotropically emitting point source orbiting this black hole to explore further signatures.
There are two reasons for us to consider the near-extremal case. First, the nice properties that the near-extremal case possess enable us to apply a powerful computational method. Second, a large number of observed supermassive black holes are thought to be rotating very fast \cite{porfyriadis2017photon}, which is the near-extremal case at least in Kerr spacetime.
The signatures we obtained have the following properties.
The sizes of shadows cast by near-extremal Kerr-MOG black holes decrease when the modified parameter is increased.
The signals produced by the orbiting hot spot are similar to those produced in a high spin Kerr (the extremal value of the reduced spin is 1) \cite{gralla2017observational}. However, the extremal Kerr-MOG black hole can have a reduced spin with a finite amount below 1 (for the cases we will consider, it ranges from 0.717 to 1) which gives a wider range of possible spin for a near-extremal astrophysical black hole. Furthermore, the observational appearance of the hot spot is also quantitatively different from Kerr. For example, when the modified parameter is increased from zero, the flux increases and the typical redshift factor for the secondary images also increases. These signatures appear periodically with period greater than that of Kerr. The critical allowed inclination for an observer to see this effect also increases in the MOG cases. This provides other possible signatures for the EHT to test.

We organize this paper as follows.
In Sec.~\ref{sec:OrbitingEmitter}, we set up the ray-tracing problem for a general Kerr-MOG black hole and write down the equations to be solved. In Sec.~\ref{sec:NearExtremalExpansion}, we solve the equations in the near-extremal limit to leading order in the deviation from extremality.
In Sec.~\ref{app:ShadowAndNHEKline}, we compute the shadow of a non-extremal Kerr-MOG black hole and discuss its near-extremal limit.
In Sec.~\ref{sec:ObservationalAppearance}, we present our results with figures and discuss these in detail. We furthermore compare our results with that of the Kerr black hole.
In App.~\ref{app:Integrals}, we introduce the computational method for the integrals that appear in our analysis and list the related results. We follow the conventions of Ref.~\cite{gralla2017observational}, but use $(x,y)$ to denote the apparent positions of the images instead of $(\alpha,\beta)$. We use $\alpha$ (and $\beta$) to describe the parameter of the MOG theory.


\section{Orbiting emitter near Kerr-MOG black hole}
\label{sec:OrbitingEmitter}

The Kerr-MOG black hole is a stationary, axisymmetric solution of the Scalar-Tensor-Vector Gravity (STVG) or modified gravity (MOG) theory \cite{moffat2006scalar}. The metric in Boyer-Lindquist coordinate reads \cite{Moffat:2014aja},
\be
ds^2=-\frac{\Delta \Sigma}{\Xi}dt^2+\frac{\Sigma}{\Delta}dr^2+\Sigma d\theta^2+\frac{\Xi \sin^2 \theta}{\Sigma}(d\phi-\omega dt)^2,
\ee
where we have set Newton's constant $G_N=1$ and defined
\begin{subequations}
\bea
\omega=\frac{a(2M_{\alpha}r-\beta^2)}{\Xi},\quad
\Delta=r^2-2M_{\alpha}r+a^2+\beta^2,\\
\Sigma=r^2+a^2\cos^2\theta,\quad
\Xi=(r^2+a^2)^2-\Delta a^2\sin^2\theta,
\eea
\end{subequations}
with
\be
\label{beta}
M_{\alpha}=(1+\alpha)M,\qquad
\beta^2=\frac{\alpha}{1+\alpha}M_{\alpha}^2,
\ee
where $M$ and $a$ are mass and spin parameters of the black hole and $\alpha$ is the deformation parameter defined by $G=G_N(1+\alpha)$ with $G$ being an enhanced gravitational constant \cite{Moffat:2014aja}. $M_{\alpha}$ and $\hat{J}=M_{\alpha}a$ are respectively the ADM mass and angular momentum of the Kerr-MOG metric \cite{sheoran2017mass}. In addition,
\be
\label{eq: gravitational charge}
K=\sqrt{\alpha G_N}M
\ee
is the gravitational charge of the MOG vector field \cite{Moffat:2014aja,sheoran2017mass}.
Solving the equation $\Delta=0$ gives radii of the inner and outer event horizons,
\be
\label{eq:EventHorizon}
r_{\pm}=M_{\alpha}\pm\sqrt{M_{\alpha}^2-(a^2+\beta^2)}.
\ee
The extremal limit is obtained for $a^2+\beta^2=M_{\alpha}^2$.

Note that the quantities under the square roots of \eqref{eq: gravitational charge} and \eqref{eq:EventHorizon} should be non-negative, thus we obtain physical bounds on the parameter $\alpha$ as \cite{sheoran2017mass}
\be
\label{eq:PhysicalBound}
0\leq\alpha\leq\frac{M_{\alpha}^2}{a^2}-1.
\ee

We assume that there exists an isotropic point source orbiting on a circular, equatorial geodesic at radius $r_s$. The angular velocity of this source is \cite{sheoran2017mass}
\be
\label{OmegaS}
\Omega_s=\pm\frac{\Gamma(r_s)}{r_s^2 \pm a\Gamma(r_s)},
\ee
where
\be
\label{Gamma}
\Gamma^2(r)=M_{\alpha}r-\beta^2,
\ee
and the plus or minus sign corresponds to direct (positive angular momentum) and retrograde orbits respectively. Here and hereafter, we use the subscript $s$ to represent ``source".
\subsection{Photon conserved quantities along trajectories}

The photon trajectories which connect a source to an observer are null geodesics in the Kerr-MOG spacetime.
There are four conserved quantities for a photon along its trajectory: the invariant mass $\mu^2=0$, the total energy $E$, the angular momentum $L$ and the Carter constant $Q$ \cite{carter1968global}. It is convenient to scale out the energy $E$ from the trajectory by introducing two rescaled quantities related to the conserved quantities $L$ and $Q$ as
\be
\label{eq:RescaledQuantities}
\hat{\lambda}=\frac{L}{E},\qquad
\hat{q}=\frac{\sqrt{Q}}{E}.
\ee
Note that the Carter constant $Q$ is non-negative for any photon passing through the equator plane and that only $\hat{q}^2$ appears in subsequent formulas, we will always have real $\hat{q}>0$ for the photons we  consider.

Using the Hamilton-Jacobi method we obtain the ray-tracing equations, which connect a source $(t_s,r_s,\theta_s,\phi_s)$ to an observer $(t_o,r_o,\theta_o,\phi_o)$, as \cite{Moffat:2014aja,carter1968global}:
\begin{subequations}
\label{eq:RayTracing}
\bea
\label{eq:RTheta}
&&-\kern-1.05em\int^{r_o}_{r_s}\frac{dr}{\pm\sqrt{\mathcal{R}(r)}}=
-\kern-1.05em\int^{\theta_o}_{\theta_s}\frac{d\theta}{\pm\sqrt{\Theta(\theta)}},\\
\label{eq:Phi}
\Delta\phi=\phi_o-\phi_s&=&-\kern-1.05em\int^{r_o}_{r_s}\frac{a}{\pm\Delta
\sqrt{\mathcal{R}(r)}}\Big(2M_{\alpha}r-\beta^2-a\hat{\lambda}\Big)dr\nn\\
&+&
-\kern-1.05em\int^{\theta_o}_{\theta_s}
\frac{\hat{\lambda}\csc^2\theta}{\pm\sqrt{\Theta(\theta)}}d\theta,\\
\label{eq:T}
\Delta t=t_o-t_s&=&-\kern-1.05em\int^{r_o}_{r_s}\frac{1}{\pm\Delta
\sqrt{\mathcal{R}(r)}}\Big[r^4+a^2\big(r^2+2M_{\alpha}r\nn\\
&-&\beta^2\big)
-a\big(
2M_{\alpha}r-\beta^2\big)\hat{\lambda}\Big]dr\nn\\
&+& -\kern-1.05em\int^{\theta_o}_{\theta_s}
\frac{a^2\cos^2\theta}{\pm\sqrt{\Theta(\theta)}}d\theta,
\eea
\end{subequations}
where
\begin{subequations}
\label{eq:Potentials}
\bea
\label{eq:DefofR}
\mathcal{R}(r)&=&\big(r^2+a^2-a\hat{\lambda}\big)^2-\Delta\Big[\hat{q}^2+\big(
a-\hat{\lambda}\big)^2\Big],\\
\label{eq:DefofTheta}
\Theta(\theta)&=&\hat{q}^2+a^2\cos^2\theta-\hat{\lambda}^2\cot^2\theta.
\eea
\end{subequations}
The function $\mathcal{R}(r)$ is called the radial ``potential" and $\Theta(\theta)$ is called the angular ``potential". $\mathcal{R}(r)=0$ corresponds to turning points in the $r$ direction and $\Theta(\theta)=0$ corresponds to turning points in the $\theta$ direction. 
Here and hereafter, the subscript $o$ stands for ``observer".

Since the integrals are to be evaluated as path integrals along a trajectory connecting the two points, we have used
the slash notation $-\kern-0.9em\int$ to distinguish them from  ordinary integrals.
The $r$ and $\theta$ turning points occur any time when the effective potential $\mathcal{R}(r)=0$ or $\Theta(\theta)=0$. The signs $\pm$ in these geodesic equations are chosen to be the same as those of $dr$ and $d\theta$ respectively, such that both LHS and RHS of Eq.~\eqref{eq:RTheta} are always positive. Given these turning points, there are different possibilities for light connecting a source to an observer. Thus we will introduce parameters $b$, $m$, $s$ to distinguish them. For the radial direction, we let $b=0$ label those direct trajectories with no turning points and $b=1$ label those reflected trajectories with one turning point. For the $\theta$ direction, we use $m\geq0$ to record the number of turning points and use $s\in\{+1,-1\}$ to denote the final sign of $p_{\theta}$ (the $\theta$-component of the photon's four-momentum). 

Since the unknowns $\phi_s$ and $t_s$ are related by $\phi_s=\Omega_s t_s$, it follows from Eqs.~\eqref{eq:Phi} and \eqref{eq:T} that
\be
\label{eq:DecoupleTs}
\Delta \phi-\Omega_s \Delta t=\phi_o-\Omega_s t_o.
\ee
We will place the observer at $\phi_o=2\pi N$ for an integer $N$ (physically equivalent to $\phi_o=0$) for all time $t_o$. Plugging $\phi_o=2\pi N$ into Eq.~\eqref{eq:DecoupleTs}, one can see that $N$ is the net winding number which records the extra windings executed by the photon relative to the emitter between its emission time and reception time \cite{gralla2017observational}. Then
the ray-tracing Eqs.\eqref{eq:RayTracing} can be re-expressed as the ``Kerr-MOG lens equations"
\begin{subequations}
\label{eq:LensEqns}
\bea
 I_r+b\tilde{I}_r&=&G^{m,s}_{\theta},\label{first eq}\\
 \! \! \! \!  J_r+b\tilde{J}_r+\frac{\hat{\lambda}G^{m,s}_{\phi}-\Omega_sa^2G^{m,s}_t}
{M_{\alpha}}&=&-\Omega_s t_o+2\pi N,\label{second eq}
\eea
\end{subequations}
where we have introduced the factor of $M_{\alpha}$ to make both equations dimensionless, and $I_r$, $\tilde{I}_r$, $J_r$, $\tilde{J}_r$ and $G^{m,s}_{i}$ ($i\in\cu{t,\theta,\phi}$) are defined in the same way as Ref.~\cite{gralla2017observational} (See also App.~\ref{app:Integrals}).

For each choice of net winding number $N\in\mathbb{Z}$, polar angular turning points $m\in\mathbb{Z}^{\ge0}$, final vertical orientation $s\in\{+1,-1\}$, and radial turning point number $b\in\{0,1\}$,
each photon trajectory is labeled by a pair of conserved quantities $(\hat{\lambda},\hat{q})$, which connects the source to an observer.
In other words, for given values of $N$, $m$, $s$, $b$, $\hat{\lambda}$ and $\hat{q}$, Eqs.~\eqref{eq:LensEqns} determine the observer coordinates $t_o$, $r_o$, $\theta_o$ for given values of the source coordinates $r_s$, $\theta_s$ (We have chosen $\phi_o=2\pi N$ and decoupled $t_s$ and $\phi_s$ using Eq.~\eqref{eq:DecoupleTs}). For a distant observer we have $r_o=\infty$ and for an equatorial source we have $\theta_s=\pi/2$. From another point of view, by solving Eq.~\eqref{eq:LensEqns} for given choice of $N$, $m$, $s$, $b$ and given values of $\theta_o$ and $r_s$, one may find the functions of $\hat{\lambda}$ and $\hat{q}$ in terms of $t_o$ which are associated with the time-dependent images of the emitter seen by the observer.


\subsection{Observational appearance}
Following Refs.~\cite{cunningham1972optical,cunningham1973optical,gralla2017observational}, we will now consider the observational appearance of the point emitter: the images positions, redshift factors and fluxes. These observational quantities can be expressed in terms of the conserved quantities.
\label{subsec:Interpretation}

The apparent position $(x,y)$ of images on the observer's screen is given by
\begin{subequations}
\label{exact position}
\bea
x&=&-\frac{\hat{\lambda}}{\sin\theta_o},\\
y&=&\pm\sqrt{\hat{q}^2+a^2\cos^2\theta_o-\hat{\lambda}^2\cot^2\theta_o}\nn\\
&&=\pm\sqrt{\Theta(\theta_o)}.
\eea
\end{subequations}
The sign $\pm$ is equal to the sign of $p_{\theta}$ (i.e., the final vertical orientation $s$) at the observer, which represents whether the photon arrives from above/below.

The ``redshift factor" $g$ is given by
\be
\label{redshift}
g=\frac{1}{\gamma}
\sqrt{\frac{\Delta_s\Sigma_s}{\Xi_s}}(1-\Omega_s \hat{\lambda}),
\ee
where we introduced the boost factor $\gamma$, which is defined as
\be
v_s=\frac{\Xi_s}{\Sigma_s\sqrt{\Delta_s}}(\Omega_s-\omega_s),\quad
\gamma=\frac{1}{\sqrt{1-v^2_s}}.
\ee

The ratio between the image flux $F_o$ and the comparable ``Newtonian flux" $F_N$ is given by
\be
\label{flux final pre}
\frac{F_o}{F_N}=g^3\frac{\hat{q}M_{\alpha}}{\gamma\sin\theta_o}
\sqrt{\frac{\Sigma_s\Delta_s}
{\Xi_s\Theta(\theta_o)\Theta(\theta_s)\mathcal{R}(r_s)}}\left|
\det\frac{\partial(B,A)}
{\partial(\hat{\lambda},\hat{q})}\right|^{-1},
\ee
where we defined
\begin{subequations}
\label{eq:DefAB}
\bea
\label{eq:DefA}
A&\equiv& I_r+b\tilde{I}_r-G^{m,s}_{\theta}\pm M_{\alpha}\int^{\theta_s}_{\pi/2}\frac{d\theta}{\sqrt{\Theta(\theta)}},\\
B&\equiv& J_r+b\tilde{J}_r+\frac{\hat{\lambda}G^{m,s}_{\phi}-\Omega_sa^2G^{m,s}_t}
{M_{\alpha}}.
\eea
\end{subequations}
The $\pm$ sign in \eqref{eq:DefA} corresponds to pushing the source above/below the equatorial plane.

Note that the conserved quantities \eqref{exact position}, \eqref{redshift} and \eqref{flux final pre} have the same form as that of Kerr \cite{gralla2017observational}, but the difference is implied via the specific expressions for $M_{\alpha}$ $\Delta_s$, $\omega_s$ and $\Omega_s$.
If we take $\alpha=0$, the above results reduce to the Kerr case (see Ref.~\cite{gralla2017observational}).
\section{Near-extremal expansion}
\label{sec:NearExtremalExpansion}

Without loss of generality, we take the observer to sit in the northern hemisphere $\theta_o\in(0, \pi/2)$ and set $M_{\alpha}=1$ in the following. We consider an emitter on, or near, the (prograde) Innermost Stable Circular Orbit (ISCO) of a near-extremal Kerr-MOG black hole. We introduce a dimensionless radial coordinate $R$ for convenience, which is related the Boyer-Lindquist radius $r$ by
\be
R=r-1,
\ee
We also introduce a small parameter $\epsilon$ to describe the deviation of the black hole from extremality,
\be
a^2+\beta^2=1-\epsilon^3\label{near extre con}
\ee
For simplicity, instead of using the parameter $\alpha$, we will use the spin $a$ in the following expressions as the free parameter that describes the modified black hole. We can get the relation between $\alpha$ and $a$ from \eqref{beta}
and \eqref{near extre con}, as
\be
\alpha=\frac{1}{a^2}-1+\mathcal{O}(\epsilon^3) \label{relation of a alpha}.
\ee
Using the standard procedure \cite{bardeen1972rotating} along with the constraint formula \eqref{near extre con}, one finds that the ISCO of a near-extremal Kerr-MOG black hole to the leading order in $\epsilon$ is at
\be
R_{\text{ISCO}}=\Big(\frac{2a^2}{2a^2-1}\Big)^{1/3}\epsilon+
\mathcal{O}(\epsilon^2).
\ee
Note that we have to choose $a>\sqrt{2}/2$ to guarantee the ISCO is in the outside of the event horizon, this corresponds to a restriction for the modified parameter $\alpha$,
\be
\label{eq:ISCOBound}
\alpha<1.
\ee

In terms of the dimensionless radius $R$, the observer is located at $R_o=(r_o-1)\approx r_o$, while the source orbits on or near the ISCO,
\be
\label{big rs}
R_s=\epsilon\bar{R}+\mathcal{O}(\epsilon^2), \qquad\bar{R}\geq\Big(\frac{2a^2}{2a^2-1}\Big)^{1/3}.
\ee
Thus, to leading order in $\epsilon$ we have
\be
\label{eq:RsExpansion}
r_s=1+\epsilon \bar{R}.
\ee
Following Ref.~\cite{gralla2017observational,porfyriadis2017photon}, we will also introduce new quantities $\lambda$ and $q$ defined by
\be
\hat{\lambda}=\frac{1+a^2}{a}(1-\epsilon \lambda),\qquad
\hat{q}=\sqrt{4-\frac{1}{a^2}-q^2}.
\ee

For later reference, we  expand the orbital frequency $\Omega_s$ and period $T_s$ in $\epsilon$, leading to the expressions
\be
\label{eq:period}
\Omega_s=\frac{a}{1+a^2}+\mathcal{O}(\epsilon),\qquad
T_s=\frac{2(1+a^2)\pi}{a}+\mathcal{O}(\epsilon).
\ee
Note that the orbital frequency/period in the near-extremal Kerr-MOG case is smaller/greater as compared to the near-extremal Kerr case \cite{gralla2017observational} for $0<a<1$.

If we take $a=1$ (corresponding to $\alpha=0$), the above results reduce to the Kerr case (see Ref.~\cite{gralla2017observational}).

\subsection{Photon conserved quantities along trajectories}
We will find solutions of Eq.~\eqref{eq:LensEqns} ($\hat{\lambda}$, $\hat{q}$, or equivalently, $\lambda$, $q$) to  leading order in $\epsilon$. Note that we must keep the $\mathcal{O}(\epsilon^0)$ terms in the equations in order to achieve a quantitative validity at reasonable values of $\epsilon$.
\subsubsection{First equation}
We start by solving the first equation~\eqref{first eq},
\be
\label{final first eq}
I_r+b\tilde{I}_r=mG_{\theta}-s\hat{G_\theta}.
\ee
The $I$ integrals and $G$ integrals are performed in App.~\ref{app:Integrals}. Since $t_o$ does not appear in the first equation, for given choice of $m$, $s$, $b$, we will express $\lambda$ as a function of $q$ by plugging the results of integrals in the equation.
Following the method of Ref.~\cite{gralla2017observational}, for each choice of $m,b,s$, and $q$, we obtain the solution of Eq.~\eqref{final first eq}
and the conditions for its existence. The conditions are given by
\begin{subequations}
\label{eq:qRange}
\bea
\bar{R}&<&\frac{4\Upsilon}{q^2}\Bigg(1+\frac{2}{\sqrt{4-q^2}}\Bigg)
\quad \text{if}\quad b=0 ,\\
\bar{R}&>&\frac{4\Upsilon}{q^2}\Bigg(1+\frac{2}{\sqrt{4-q^2}}\Bigg)
\quad \text{if}\quad b=1 ,
\eea
\end{subequations}
and the solution is
\be
\lambda=\frac{4\Upsilon}{(1+a^2)(4-q^2)}\Bigg[2- q\sqrt{1+\frac{4-q^2}{2\Upsilon}\bar{R}}\Bigg] \ . \label{lambda of q}
\ee
Here, $\Upsilon >0$ is defined by
\be
\Upsilon\equiv\frac{q^4R_o}{q^2+2R_o+qD_o}e^{-qG^{\bar{m},s}_{\theta}}=
\frac{q^4}{q+2}e^{-qG^{\bar{m},s}_{\theta}}+\mathcal{O}(\frac{1}{R_o}),
\ee
where $D_o$ is defined in \eqref{eq:Do} and $G^{\bar{m},s}_{\theta}$ is defined in \eqref{eq:AngleIntegralMS} and \eqref{eq:AngleIntegral}, with
\be
\bar{m}=m+\frac{1}{qG_{\theta}}\log\epsilon.
\ee
Note that $\Upsilon$ is independent of $R_o$ for large $R_o$.

\subsubsection{Second equation}
Next we move on to second equation \eqref{second eq} which gives another relation between $t_o$, $\lambda$ and $q$ for given choice of $m$, $s$, $b$. We will look for functions $\lambda(\hat{t}_o)$ and $q(\hat{t}_o)$ which are associated with the time-dependent tracks of images. We introduce a dimensionless time coordinate $\hat{t}_o$ such that the emitter has unit periodicity in terms of it,
\be
\hat{t}_o=\frac{t_o}{T_s}=\frac{at_o}{2(1+a^2)\pi}+\mathcal{O}(\epsilon).
\ee
Eq.~\eqref{second eq} can be rewritten in terms of this dimensionless time coordinate, as
\begin{subequations}
\label{definition G}
\bea
\hat{t}_o&=&N+\mathcal{G},\\
\mathcal{G}&\equiv&-\frac{1}{2\pi}\Big(J_r+b\tilde{J}_r+\frac{1+a^2}{a}G^{m,s}_{\phi}
-\frac{a^3 G^{m,s}_t}{1+a^2}\Big).
\eea
\end{subequations}
The $J$ integrals and $G$ integrals are given in App.~\ref{app:Integrals}.

Since the problem is periodic, we will consider the single period $\hat{t}_o\in[0,1]$. For each given choice of $m$, $s$, $b$ having a non-zero range of $q$ satisfying the condition \eqref{eq:qRange}, the first equation \eqref{first eq} gives a function $\lambda(q)$ (\eqref{lambda of q}), the second equation \eqref{second eq} (or equivalently, Eq.~\eqref{definition G}) then gives a function $\hat{t}_o(q)$ for each choice of an integer $N$. In the given period $0\leq\hat{t}_o<1$, $N$ is uniquely determined for each $q$, and the multivalued inverse $q(\hat{t}_o)$ corresponds to the time-dependent tracks of images. For each allowed $N$ within the corresponding range of $q$, $\hat{t}_o(q)$ may either be monotonic or has local maxima and/or minima. 
For the monotonic ones, we are able to get the inverse $q(\hat{t}_o)$. For the non-monotonic ones, we divide $\hat{t}_o(q)$ into several invertible parts to get their inverse and label each inverse $q_i(\hat{t}_o)$ with a discrete integer $i$. The image track can then be divided into several segments which associate with these functions $q(\hat{t}_o)$, each such track segment is uniquely labeled by $(m,b,s,N,i)$. Finding all the track segments for all choices of $m$, $b$, $s$, $N$, $i$ gives the tracks of all the images which associate with the complete observable information. We give an example in Sec.~\ref{sec:ObservationalAppearance} to describe a practical approach.

\subsubsection{Winding number around the axis of symmetry}
The winding number around the axis of symmetry for a photon trajectory is $n=\text{mod}_{2\pi}\Delta\phi$, where $\Delta\phi$ can be obtained from Eq.~\eqref{eq:Phi}. Using the method of matched asymptotic expansions (MAE) described in App.~\ref{app:MAEandRadialIntegral}, $\Delta\phi$ is expressed to the leading order in $\epsilon$ as
\be
\Delta\phi=\frac{a}{(1+a^2)\lambda \epsilon}\bigg(\frac{D_s}{\bar{R}}-q\bigg)
+\mathcal{O}(\log\epsilon),
\ee
where $D_s$ is defined in \eqref{eq:Ds}. Note that this leading order expression scales as $\epsilon^{-1}$.

\subsection{Observational appearance}
Recall from the beginning of Sec.~\ref{sec:NearExtremalExpansion} that for near-extremal Kerr-MOG we have
\begin{subequations}
\label{eq:RecallExpansion}
\bea
\alpha&=&\frac{1}{a^2}-1+\mathcal{O}(\epsilon^3),\qquad
r_s=1+\epsilon\bar{R},\\
\hat{\lambda}&=&\frac{1+a^2}{a}(1+\epsilon\lambda),\qquad
\hat{q}=\sqrt{4-\frac{1}{a^2}-q^2}.
\eea
\end{subequations}
We now expand the observational quantities for an individual image to the leading order in $\epsilon$. The quantities involved
are  the position \eqref{exact position}, the redshift factor \eqref{redshift} and the flux \eqref{flux final pre}.
\subsubsection{Image positions and redshift factors}
The image position \eqref{exact position} on the observer's screen is expanded as
\begin{subequations}
\label{image position}
\bea
x&=&-\frac{1+a^2}{a}\frac{1}{\sin\theta_o}+\mathcal{O}(\epsilon),\\
y&=&s\sqrt{4-\frac{1}{a^2}-q^2+a^2\cos^2\theta_o-\frac{(1+a^2)^2}{a^2}\cot^2
\theta_o}\nn\\&&+\mathcal{O}(\epsilon).
\eea
\end{subequations}
Note that the leading order of position does not include $\lambda$. We should impose a requirement that $y$ is real to make sure the photons can reach infinity, which gives a range of $q$:
\be
\label{range of q}
q\in\Bigg[0,\sqrt{4-\frac{1}{a^2}+a^2\cos^2\theta_o-\frac{(1+a^2)^2}{a^2}\cot^2
\theta_o}\Bigg].
\ee
As $\epsilon\rightarrow0$, Eqs.~\eqref{image position} and \eqref{range of q} gives a vertical line segment on which all images of the hot spot appear. We call this vertical line the NHEK-MOG line, being  the analog of NHEKline for Kerr \cite{gralla2017observational} (Sec.~\ref{app:ShadowAndNHEKline}). We find that there is no range of $q$ at all when $\theta_o<\theta_{\text{crit}}$, which means that the NHEK-MOG line will disappear (Sec.~\ref{app:ShadowAndNHEKline}) in that case.

The redshift \eqref{redshift} is expanded as
\be
\label{redcorr}
g=\frac{1}{\frac{\sqrt{4a^2-1}}{a}+\frac{2a(1+a^2)}
{\sqrt{4a^2-1}}\frac{\lambda}{\bar{R}}}+\mathcal{O}(\epsilon).
\ee
Note that the cosines of emissive direction establish an above bound for $g$ \cite{gralla2017observational}, which is
\be
g\leq \frac{a(1+2a)}{\sqrt{4a^2-1}}.
\ee
\subsubsection{Image fluxes}
The ratio of image flux \eqref{flux final pre} is expanded as
\bea
\label{flux extremal}
\frac{F_o}{F_N}&=&\frac{\sqrt{4a^2-1}\epsilon\bar{R}}{2a^2D_s}\frac{qg^3}
{\sqrt{4-
\frac{1}{a^2}-q^2}\sqrt{\Theta_0(\theta_o)}\sin\theta_o}\nn\\
&&\times\left|\det\frac{\partial(B,A)}{\partial(\lambda,q)}\right|^{-1},
\eea
where $g$ and $D_s$ are given in Eqs.~\eqref{redcorr} and \eqref{eq:Ds}, and (see Eq. \eqref{image position})
\bea
\Theta_0(\theta_o)=\Theta(\theta_o)\big|_{\lambda=0}&=&
4-\frac{1}{a^2}-q^2+a^2\cos^2\theta_o\nn\\
&-&\frac{(1+a^2)^2}{a^2}\cot^2
\theta_o=y^2,
\eea
and (recall the definition of $A$ and $B$, the Eq.~\eqref{eq:DefAB})
\bea
\label{eq:JacabiBLambda}
\left|\det\frac{\partial (B,A)}{\partial (\lambda,q)}\right|&=&
\bigg|\frac{\partial}{\partial \lambda}\bigg(J_r+b\tilde{J}_r\bigg)\bigg[\frac{\partial}{\partial q}\bigg(I_r+b\tilde{I}_r\bigg)-\frac{\partial G^{m,s}_{\theta}}{\partial q}\bigg]\nn\\
&-&\frac{\partial}{\partial\lambda}\bigg(I_r+b\tilde{I}_r\bigg)\bigg[
\frac{\partial}{\partial q}\bigg(J_r+b\tilde{J}_r\bigg)+\frac{\partial G^{m,s}_{t\phi}}{\partial q}\bigg]\bigg|\nn\\
&+&\mathcal{O}(\epsilon\log\epsilon),
\eea
where we introduced
\bea
G^{m,s}_{t\phi}&=&\hat{\lambda}G^{m,s}_{\phi}-\Omega_s a^2G^{m,s}_t\nn\\
&=&\frac{1+a^2}{a}G^{m,s}_{\phi}-\frac{a^3}{1+a^2}G^{m,s}_t+\mathcal{O}(\epsilon).
\eea
The $G$, $I$, $J$ integrals and variations of $I$, $J$ integrals are given in App.~\ref{app:Integrals}.

\section{Shadow and NHEK-MOG line}
\label{app:ShadowAndNHEKline}
The entire image of a black hole seen from the Event Horizon Telescope is expected to be the black hole ``shadow". To understand the images of the emitter better, we compute the edge of a shadow cast by a near-extremal Kerr-MOG black hole.
The edge of a black hole shadow is the boundary from where the photons can escape \cite{bardeen1973timelike}, which corresponds to spherical massless geodesics with fixed $r=\tilde{r}$.

First, we consider the non-extremal case and restore $M_\alpha$. These ``spherical photon orbits" satisfy
\be
\label{spherical con}
\mathcal{R}(\tilde{r})=\mathcal{R}^{\prime}(\tilde{r})=0,
\ee
where $\mathcal{R}(r)$ is defined in \eqref{eq:DefofR} and the prime represents derivative. For a Kerr-MOG black hole we have $0<a^2+\beta^2<M_{\alpha}^2$. Then from Eq.~\eqref{spherical con} we get
\begin{subequations}
\label{exactLambdaQ}
\bea
\hat{\lambda}&=&-\frac{\tilde{r}(\tilde{r}^2-M_{\alpha}\tilde{r}
-2\Gamma(\tilde{r})^2)+a^2(\tilde{r}+M_{\alpha})}{a(\tilde{r}-M_{\alpha})},\\
\hat{q}&=&\frac{\tilde{r}\sqrt{4a^2\Gamma(\tilde{r})^2-(\tilde{r}^2-
M_{\alpha}\tilde{r}-2\Gamma(\tilde{r})^2)^2}}{a(\tilde{r}-M_{\alpha})},
\eea
\end{subequations}
where $\Gamma(\tilde{r})$ is defined in \eqref{Gamma}.
The shadow edge is the curve $(x(\tilde{r}),y(\tilde{r}))$ obtained by substituting Eqs. \eqref{exactLambdaQ} for $\hat{\lambda}$ and $\hat{q}$ into Eq. \eqref{exact position}.
\subsection{Extremal limit}
Following the procedure of Ref.~\cite{gralla2017observational}, we now consider the extremal limit. We set $M_{\alpha}=1$ again and let $a^2+\beta^2\rightarrow1$. Then Eq.~\eqref{exactLambdaQ} gives
\begin{subequations}
\bea
\hat{\lambda}&=&-\frac{1}{a}(\tilde{r}^2-2\tilde{r}-a^2),\\
\hat{q}&=&\frac{\tilde{r}}{a}\sqrt{4a^2-(\tilde{r}-2)^2},
\eea
\end{subequations}
and the condition $\hat{q}$ is real gives the range of $\tilde{r}$
\be
\tilde{r}\in[\tilde{r}_-,\tilde{r}_+]=[1,2(a+1)],
\ee
The shadow edge is then given by the curve
\begin{subequations}
\label{eq:OpenPosition}
\bea
x(\tilde{r})&=&\frac{1}{a}(\tilde{r}^2-2\tilde{r}-a^2)\csc\theta_o,\label{open x}\\
y(\tilde{r})&=&\pm\Bigg[\frac{\tilde{r}^2}{a^2}\big(4a^2-(\tilde{r}-
2)^2\big)+a^2\cos^2\theta_o\nn\\
&&-\Big(\frac{\tilde{r}^2-2\tilde{r}-a^2}{a}
\Big)^2\cot^2\theta_o\Bigg]^{1/2}.\label{open y}
\eea
\end{subequations}
However, curves given by these equations are not closed in general. We show the curves for different values of $a$ in the dashed line in Fig.~\ref{fig:Shadow}. The two endpoints are at the positions where $\tilde{r}=\tilde{r}_-=1$, which are given by
\begin{subequations}
\bea
x_{\text{end}}&=&-\frac{1+a^2}{a}\csc\theta_o, \\
y_{\text{end}}&=&\pm\sqrt{4-\frac{1}{a^2}+a^2\cos^2\theta_o-\frac{(1+a^2)^2}
{a^2}\cot^2\theta_o}.
\eea
\end{subequations}
Note that there are no endpoints at all when the quantity under the square root is negative, and thus the curve is closed. The condition for an open curve is $\theta_{\text{crit}}<\theta_o<\pi-\theta_{\text{crit}}$, where
\be
\label{eq:Inclination}
\theta_{\text{crit}}=\arctan\sqrt{\frac{a^2+3-2\sqrt{2+a^2}}{2\sqrt{2+a^2}-3}}.
\ee

Therefore, this $a^2+\beta^2\rightarrow1$ limit has missed an important piece for the open curve. To make the curve be closed in that case, we reconsider this limit by introducing an alternative parameter $\tilde{R}$ defined by
\be
a^2+\beta^2=1-\delta^2,\qquad
\tilde{r}=1+\delta\tilde{R}.
\ee
As $\delta\rightarrow0$, we recover the missing part of the shadow, which is given by
\begin{subequations}
\bea
\label{NHEK}
x&=&-\frac{1+a^2}{a}\csc\theta_o,\\
|y|&<&\sqrt{4-\frac{1}{a^2}+a^2\cos^2\theta_o-\frac{(1+a^2)^2}
{a^2}\cot^2\theta_o}.
\eea
\end{subequations}
We show this in the solid lines in Fig.~\ref{fig:Shadow} and name this segment the NHEK-MOG line which is the analog of the NHEKline in Kerr spacetime \cite{gralla2017observational}. We find the images of an  emitter orbiting on (or near) the ISCO appear on the NHEK-MOG line (see from Eq.~\eqref{image position}).

Therefore, the extremal limit of the Kerr-MOG black hole shadow is given by the union of the open curve \eqref{eq:OpenPosition} and the NHEK-MOG line \eqref{NHEK}. 

\section{Results and discussion}
\label{sec:ObservationalAppearance}
\begin{table}[ht!]
\centering
 \begin{tabular}{|l|l|l|l|l|}
        \hline
         No.  & a & $\alpha$ &$ \bar{R}_{\text{ISCO}}$& $\theta_{\text{crit}}$ \\
        \hline
          1 & 0.717 & 0.945  & 3.317& 54.758$^\circ$ \\
        \hline
          2& 0.75& 0.778&2.080&53.228$^\circ$ \\
          \hline
          3&0.8&0.563&1.660&51.353$^\circ$ \\
          \hline
          4&0.85&0.384&1.481&49.881$^\circ$ \\
          \hline
          5&0.9&0.235&1.377&48.716$^\circ$ \\
          \hline
          6&0.95&0.108&1.309&47.792$^\circ$ \\
          \hline
          7&1&0&1.260&47.059$^\circ$ \\
        \hline
       \end{tabular}
        \caption{The range of the deformation parameters $\alpha$, the dimensionless radii of the ISCO $\bar{R}_{\text{ISCO}}$ (Eq.~\eqref{eq:parameters}) and the critical observer inclinations $\theta_{\text{crit}}$ (Eq.~\eqref{eq:Inclination}), corresponding to different values of the spin parameters $a$ for the extremal Kerr-MOG black holes, where we choose $a=\sqrt{2}/2+10^{-2}\approx0.717$ as the critical case.}
        \label{table 1}
  \end{table}

We now describe the results with figures and discuss them in detail. First we will look at the whole silhouettes (shadow) of a near-extremal Kerr-MOG black hole and discuss how the size is changed when the free parameter $\alpha$ (we will also equivalently use $a$ as the free parameter in later discussion since there is the relation \eqref{relation of a alpha} between them for the near-extremal cases) is changed. Then, we will focus on a special portion of the shadow, which we call the ``NHEK-MOG line" (in analogy to the so-called NHEKline in Ref.~\cite{gralla2017observational}), where the images of the point emitter appear. These images have some characteristic features which are similar to that of a near-extremal (high spin) Kerr black hole.

The modified parameter that we wish to consider should satisfy the physical bounds \eqref{eq:PhysicalBound} and \eqref{eq:ISCOBound}, which gives a range of $0\leq\alpha<1$ (corresponding to $1\geq a>\sqrt{2}/2$). These choices are also in the allowed range for supermassive black holes, $0.03<\alpha<2.47$ \cite{Perez:2017spz} (except for the critical case $\alpha=0$).
For each choice of the modified parameter $\alpha$ (in our formulae we use $a$ instead), the observable quantities of a hot spot depend on four parameters, $\epsilon$, $\bar{R}$, $R_o$, and $\theta_o$. To make our approximations sufficiently accurate, one must choose $\epsilon\ll1$ and $R_o\gg1$ . For the emitter to be on a stable orbit of a near-extremal Kerr-MOG black hole, one must choose $\bar{R}\geq\big(\frac{2a^2}{2a^2-1}\big)^{1/3}$. To ensure that an observer can possibly see the flux, one needs to set the observer on a place with inclination satisfying $\arctan\sqrt{\frac{a^2+3-2\sqrt{2+a^2}}{2\sqrt{2+a^2}-3}}
<\theta_o<\frac{\pi}{2}$. When $a=1$ it reduces to the Kerr case. We will consider a special example with the same choice of parameters as in \cite{gralla2017observational}, in order to compare the results. The parameters are as follows:
\begin{subequations}
\bea
R_o&=&100,\qquad
\theta_o=\frac{\pi}{2}-\frac{1}{10}=84.27^\circ,\\
\epsilon&=&10^{-2},
\qquad \bar{R}=\bar{R}_{\text{ISCO}}=\Big(\frac{2a^2}{2a^2-1}\Big)^{1/3}.\label{eq:parameters}
\eea
\end{subequations}
This describes an emitter (or hot spot) on the ISCO of a near-extremal Kerr-MOG black hole with spin $a$, viewed from a nearly edge-on inclination. (Note that the parameter $a$ is the spin of a precisely extremal black hole, however, it is also the spin of a near-extremal black hole to leading order in $\epsilon$. Here and hereafter, we ignore this difference.) Table~\ref{table 1} shows the ranges of $\alpha$, $\bar{R}_{\text{ISCO}}$ and $\theta_{\text{crit}}$ corresponding to different values of $a$.  We find that $\bar{R}_{\text{ISCO}}$ and $\theta_{\text{crit}}$ (\eqref{eq:Inclination}) increase when $\alpha$ is increased and that $a$ decreases when $\alpha$ is increased. This agrees with Ref.~\cite{lee2017innermost} where the authors find that the ISCOs of Kerr-MOG black holes are always greater than that of Kerr black hole.

\subsection{Silhouettes of black hole}
\label{subsec:Silhouetes}
\begin{figure}[ht!]
	
\begin{center}
\includegraphics[width=8cm]{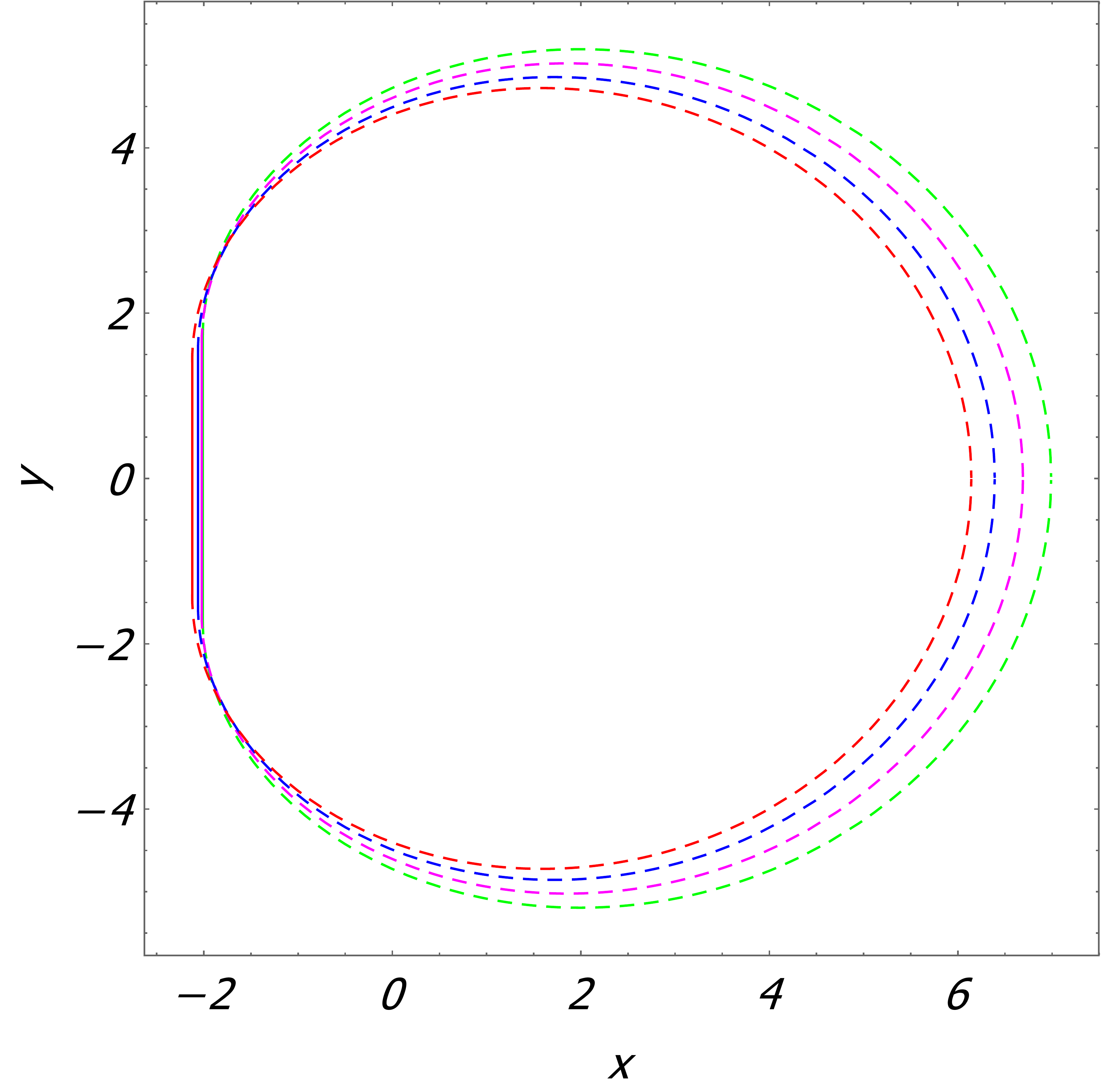}
\end{center}
	\caption{Edges of near-extremal Kerr-MOG black hole shadows, where the dashed lines are the open parts \eqref{eq:OpenPosition} and the vertical solid lines are the NHEK-MOG lines \eqref{NHEK}. The green, magenta, blue and red curves have $a=1$, $0.9$, $0.8$, $0.717$ respectively.}
	\label{fig:Shadow}
\end{figure}

Fig.~\ref{fig:Shadow} shows the edges of near-extremal Kerr-MOG black hole shadows (see Sec.~\ref{app:ShadowAndNHEKline}). We find that the sizes of shadows cast by a near-extremal Kerr-MOG black hole decrease when the free parameter $\alpha$ is increased from zero. The length of the NHEK-MOG line and the angle corresponding to it also decrease while the free parameter $\alpha$ is increased from zero.
In Ref.~\cite{Moffat:2015kva}, Moffat found that the sizes of shadows cast by Kerr-MOG black holes increase significantly as the free parameter $\alpha$ is increased from zero. This is not conflicting with our results because in that paper one compares the sizes of shadows for black holes with same parameter $M$, while we compare that for black holes with same ADM mass $M_{\alpha}=(1+\alpha)M$.

\subsection{Images on the NHEK-MOG line}
\label{subsec:Images}

Following the procedure of Ref.~\cite{gralla2017observational} and using the open numerical code therein, we now show the images of the of an emitter orbiting on ISCO.
As discussed in Sec.~\ref{app:ShadowAndNHEKline}, the images appear on a vertical line, the NHEK-MOG line. These images are photons arriving with different combinations of the discrete parameters $m$, $s$, $b$ $N$ as well as an additional label $i$ if the function $\hat{t}_o(q)$ (or equivalently $\mathcal{G}(q)$) has maxima or minima (see the discussion below Eq.~\eqref{definition G}). To show this, we choose $m$, $s$, $b$ first and then find the allowed range of $N$ as well as the possible $i$ for those have extra images due to maxima or minima.
Fig.~\ref{Fig 2} shows examples of such segments for $a=1$ and $a=0.8$. We choose $m=2$, $b=0$, $s=+1$ with parameter choices of \eqref{eq:parameters}, which is same as Ref. \cite{gralla2017observational} for comparison.
We find that the cases with $a=1$ (the Kerr case) and $a=0.8$ have similar features but the modified parameter $\alpha$ causes corrections to the associated functions: $\mathcal{G}$ decreases and $F_o/F_N$ increases while $a$ is decreased.

For each track segment $q(\hat{t}_o)$ labeled by $(m,b,s,N,i)$, we may determine $\lambda(\hat{t}_o)$ by Eq.~\eqref{lambda of q}. From these two conserved quantities we may then compute the main observables for the segment. The observables involved are the image position $(x,y)$ (Eq.\eqref{image position}), the image redshift $g$ (Eq.~\eqref{redcorr}), and the image flux $F_o/F_N$ (Eq.~\eqref{flux extremal}).
We then build up the complete observable information by including all such track segments. Note that only a few values of $N$ and $m$ are important because the flux of others are vanishingly small (see Fig.~\ref{Fig 2} and Fig.~\ref{fig:PeakMvalue} for details).
Below we  describe the most important features of the images in Fig.~\ref{fig:PositionFluxRedshift}.

\begin{widetext}{2}
\begin{figure}[ht!]
 \begin{center}
\includegraphics[width=\textwidth]{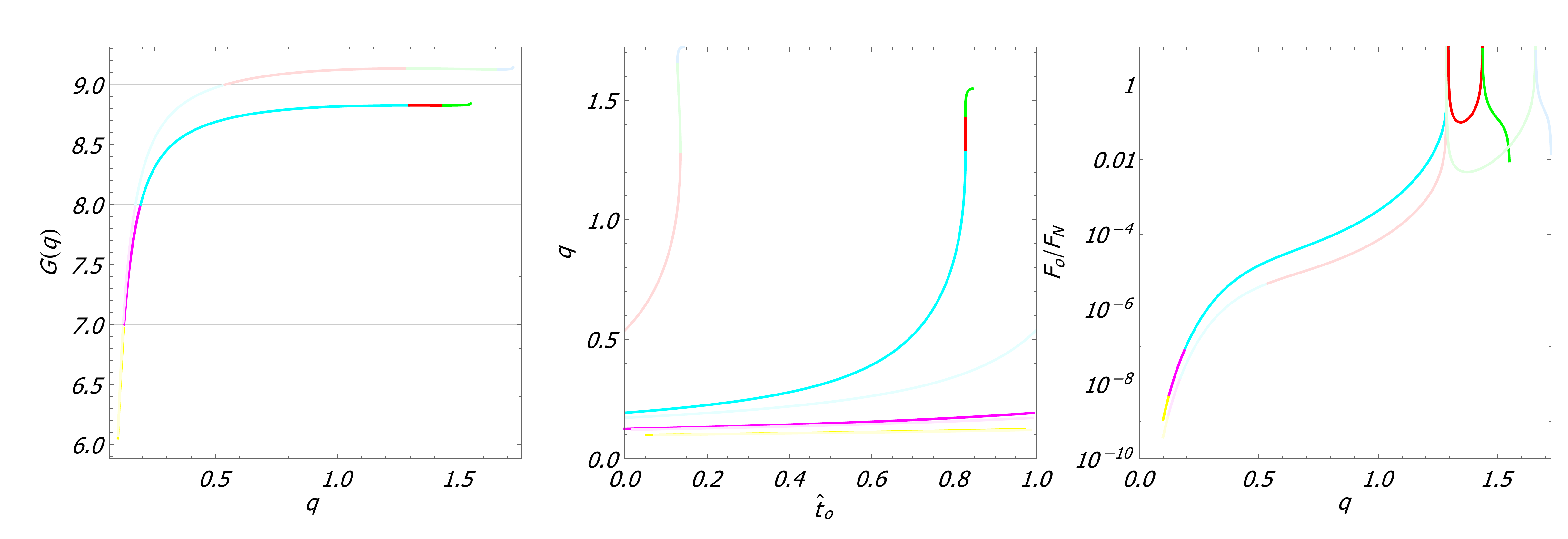}
\end{center}
  \caption{Left to right: plots of $\mathcal{G}(q)$, $q(\hat{t}_o)$ and $F_o/F_N (q)$ for $a=1$ (light curves) and $a=0.8$ (bright curves) with $m=2$, $b=0$, $s=+1$ and the parameters choices of \eqref{eq:parameters}. For $a=1$, the light yellow, light magenta and light cyan curves have $N=-6,-7,-8$, respectively, and no extra label $i$, while the light red, light green and light blue curves have $N=-9$ and $i=1,2,3$, respectively. For $a=0.8$, the yellow and magenta curves have $N=-6,-7$, respectively, and no extra label $i$, while the cyan, red and green curves have $N=-8$ and $i=1,2,3$, respectively. Note that for $a=1$, the Kerr-MOG case reduce to the Kerr case so that the light curves agree with those in Ref.~\cite{gralla2017observational} exactly. (Although the condition \eqref{eq:qRange} allows the whole range of $q$, we have imposed a small $q$ cutoff since the corresponding image fluxes are negligibly small.)}
  \label{Fig 2}
\end{figure}
\end{widetext}

Fig.~\ref{fig:PositionFluxRedshift} shows the main observables for three different values of spin, $a=1$, $a=0.8$ and $a=0.717$, with the small parameters choose as $\epsilon=0.01$. In each case, the green line is a bright primary image while others are secondary images. For $a=1$, it reduces to the Kerr case and we see that our results exactly agree with Ref.~\cite{gralla2017observational}. For $a=0.8$ and $a=0.717$ (the critical case), the general features are qualitatively similar to the case of $a=1$ but quantitatively corrected. The track segments line up in to continues tracks and flash \cite{gralla2017observational} when different tracks intersect, and the typical positions of images do not change. However,
from Fig.~\ref{fig:Shadow} we see that the length of the NHEK-MOG line decreases when $a$ is decreased, i.e. the maximum value of the screen coordinate $y_{\text{max}}$ decreases when the free parameter $\alpha$ is increased from zero.
The flux intensity increases when the black hole spin is decreased. This is also true in the near-extremal Kerr cases when considering different values of $\epsilon$ (and considering the precise spins of the black holes) \cite{gralla2017observational} since the typical flux scales as $\epsilon/\log\epsilon$. In each case for $a=1$, $0.8$, $0.717$, the primary image appears near the center of the NHEK-MOG line before moving downward while blueshifting and spiking in brightness. The winding number of these segments of primary images in the modified cases are decreased when $\alpha$ is increased. For example, the winding numbers range between 17 and 23, 11 and 16, 6 and 8, for spin $a=1$, $0.8$, and $0.717$, respectively (see Fig.~\ref{fig:PeakFlux}). For the near-extremal Kerr-MOG cases, the peak redshift factors are all at $g\approx1.6$ but correspond to different emission angles. For example, they correspond to light emitted in cones of $27^{\circ}$, $20^{\circ}$ and $25^{\circ}$ around the forward direction \cite{gralla2017observational} for $a=1$, $0.8$ and $0.717$, respectively. However, another typical redshift factor (corresponding to $\lambda\sim0$) associated with the secondary images increases when the spin parameter $a$ is decreased, at $g=a/\sqrt{4a^2-1}$.  For the near-extremal Kerr cases, both typical redshifts do not change when the spin is increased (by choosing different value of $\epsilon$ and considering the precise spins of the black holes )\cite{gralla2017observational}. The reason for this difference is that the range of spin is very limited for the near-extremal Kerr black hole, but it becomes wider for near-extremal Kerr-MOG black holes. This is why we only take $\epsilon=0.01$ into consideration. In addition, the above signatures in the near-extremal Kerr-MOG case appear periodically with period greater than that in near-extremal Kerr case (see Eq.~\eqref{eq:period}).

The typical redshift factors are related to observations since they could shift the iron line $E_{\text{FeK$\alpha$}}=6.4$ keV to $6.4g$ keV. For blueshifted primary image, the factor $g\approx1.6$ will shift the iron line to 10.2 keV. For redshifted secondary images, the redshift factors for $a=1$, $0.8$ and $0.717$ will shift the iron to 3.7 keV, 4.1 keV and 4.5 keV respectively.  However, rather than close to the observed peak at 3.5 keV \cite{carlson20153}, they are even more away from it in modified cases than in Kerr case \cite{gralla2017observational}.

\begin{widetext}{2}
\begin{figure}[ht!]	
\begin{center}
\includegraphics[width=\textwidth]{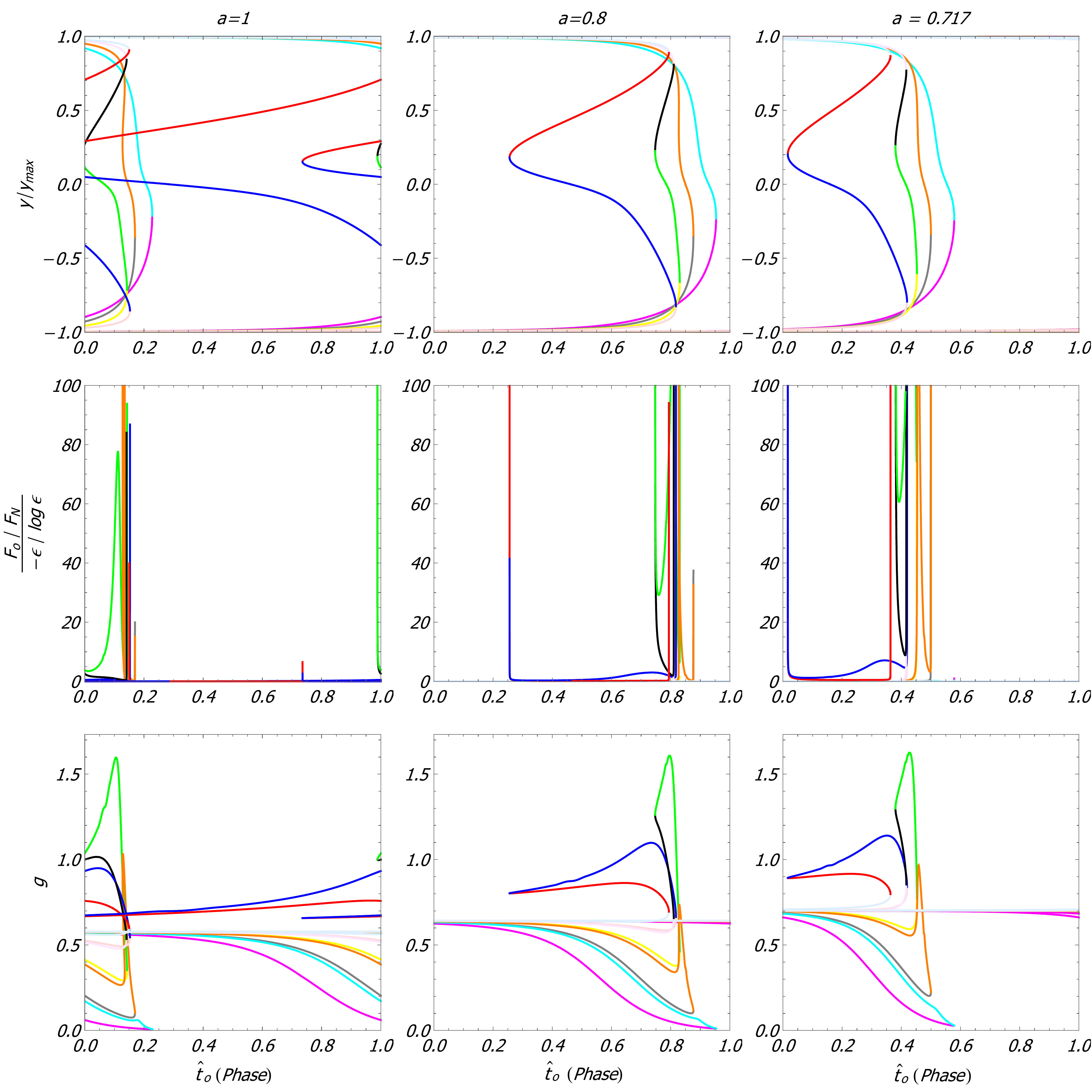}
\end{center}
	\caption{Observables of the most important few images for three different values of near-extremal spin of Kerr-MOG black holes with the parameter choices of \eqref{eq:parameters}. From top to bottom, we plot positions, fluxes and redshift factors. Form left to right, we have $a=1$ (Kerr case~\cite{gralla2017observational}), $a=0.8$ and $a=0.717$ (critical case for there exist ISCO for a near-extremal Kerr-MOG black hole). The color-coding is the same as that of Ref.~\cite{gralla2017observational}: each of these colored lines maybe a composed of continuous multiple track segments. For example, the green line (denoting the primary image) is composed of 4, 3, and 3 segments in the $a=1$, 0.8, 0.717 cases respectively.}
	\label{fig:PositionFluxRedshift}
\end{figure}

\begin{figure}[ht!]	
\begin{center}
\includegraphics[width=\textwidth]{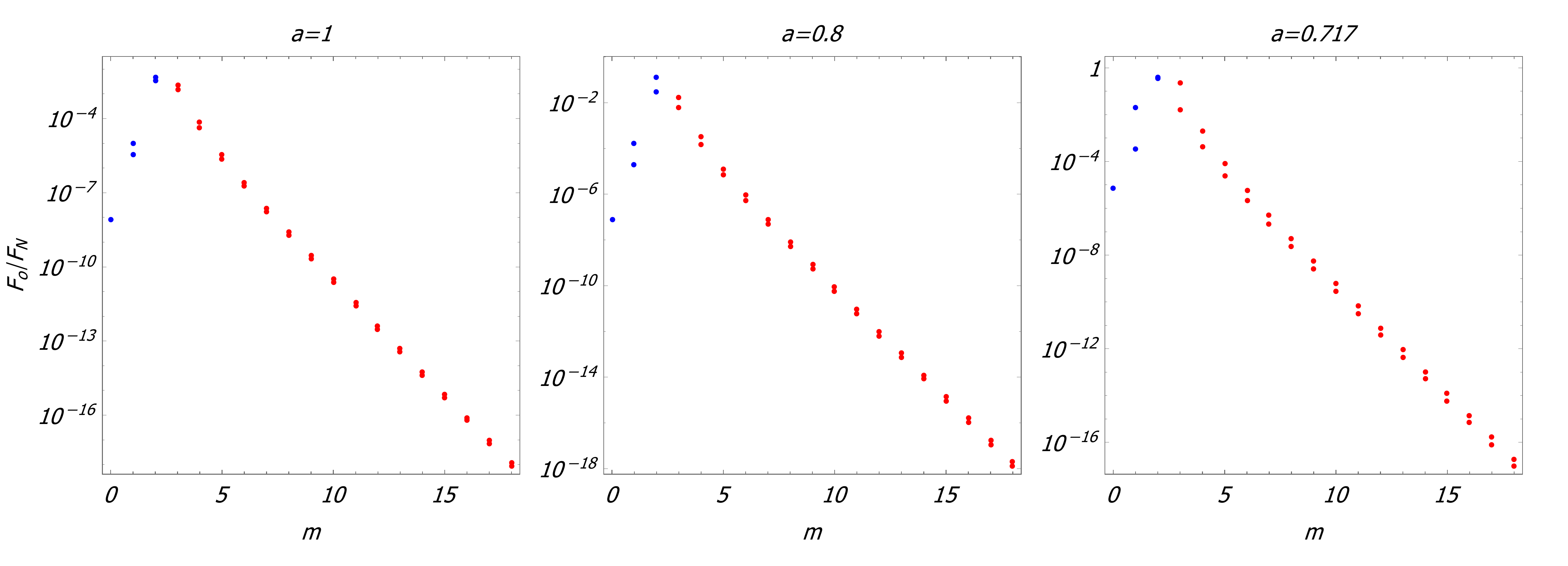}
\end{center}
	\caption{Left to right: plots of $F_o/F_N$ for $a=1$ (Kerr case~\cite{gralla2017observational}), $a=0.8$ and $a=0.717$ (critical case for there exist ISCO for a near-extremal Kerr-MOG black hole) with parameter choices of \eqref{eq:parameters}. We set $q=1.5$, $1.38$, and $1.35$ for $a=1$, $0.8$, and $0.717$, respectively, and we let $m$ vary in each case. We denote the direct/reflect (b=0/b=1) images by blue/red dots. For $m=0$, we have only one image corresponding to $s=-1$, for each other value of $m$, we have two images corresponding to $s=\pm1$. }
	\label{fig:PeakMvalue}
\end{figure}

\begin{figure}[ht!]	
\begin{center}
\includegraphics[width=\textwidth]{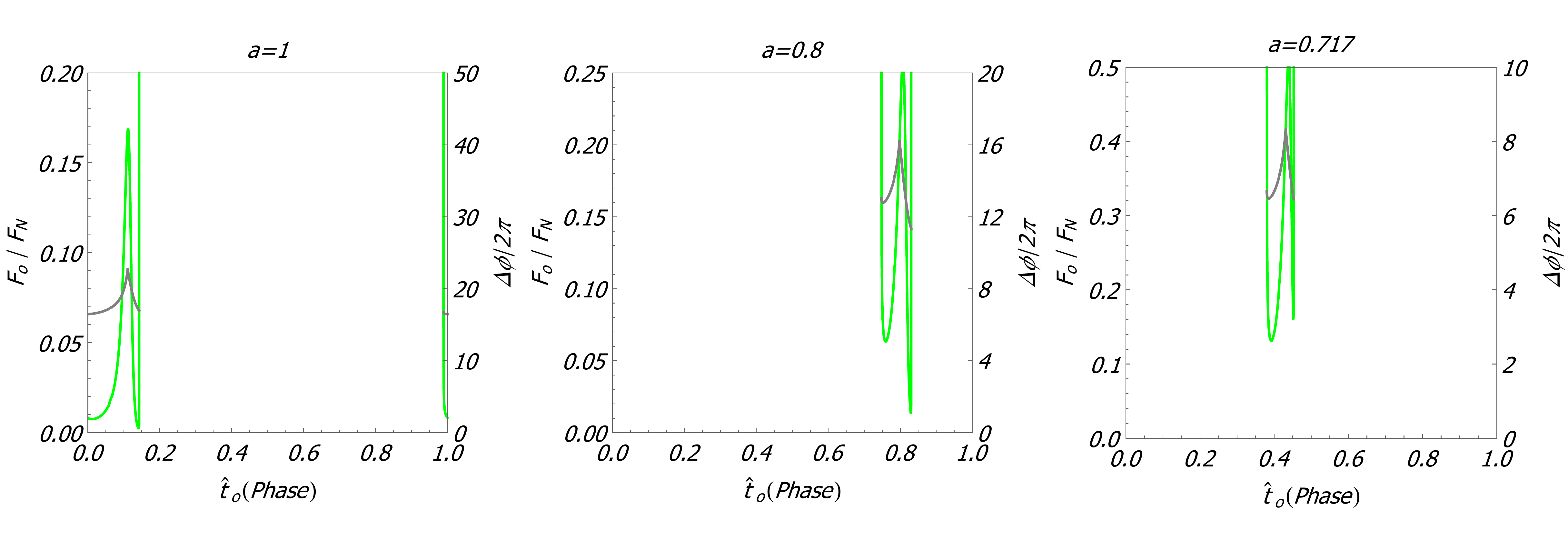}
\end{center}
	\caption{Left to right: plots of flux $F_o/F_N$ (green) and winding number $\Delta\phi/2\pi$ (gray) of the primary image for $a=1$ (Kerr case~\cite{gralla2017observational}), $a=0.8$ and $a=0.717$ (critical case for there exist ISCO for a near-extremal Kerr-MOG black hole) with parameter choices of \eqref{eq:parameters}. Note that as mentioned below Fig.~\ref{fig:PositionFluxRedshift}, the single primary image for each case is a composed of multiple track segments.}
	\label{fig:PeakFlux}
\end{figure}
\end{widetext}

\section{Summary}
In this paper, we analytically compute the observational signature of a near-extremal rotating black hole in the modified gravity theory (MOG), which is also referred as scalar-tensor-vector theory (STVG). The rotating black hole in this theory called as the Kerr-MOG black hole, introducing a modified parameter $\alpha$ in addition to the parameters of Kerr black hole. When the parameter $\alpha$ goes to zero, the modified black hole reduces to Kerr black hole. The range of the modified parameter that we considered is $0\leq\alpha<1$, which is in the supposed range for a supermassive black hole \cite{Perez:2017spz}. To be specific, we compute the near-extremal Kerr-MOG black hole's shadow and the position, redshift and flux of a orbiting hot spot's image. Compared with the signature produced in the Kerr background, the MOG case exhibits the following differences:
\begin{enumerate}
  \item The size of the shadow cast by a Kerr-MOG black hole decreases when the modified parameter $\alpha$ is increased.
  \item  The targeted astrophysical black hole could be one that has a smaller spin than a corresponding near-extremal Kerr black hole, since the spin parameter $a$ can be in the range $0.717 < a < 1$. The spin of a near-extremal Kerr-MOG black hole decreases when the modified parameter is increased.
  \item The image of the hot spot appears in the leftmost vertical line (NHEK-MOG line) of the shadow periodically with a period greater than that of Kerr. The period increases when the modified parameter is increased.
  \item The flux of the image increases when the modified parameter is increased.
  \item The typical redshift associated with the secondary image increases when the modified parameter is increased.
\end{enumerate}

\subsection*{Acknowledgements}
We are grateful for the open numerical code of Ref.~\cite{gralla2017observational}. We thank Jay Armas and Maria J. Rodriguez for reading the manuscript of this paper and for useful comments. We thank John W. Moffat for helping us understand the MOG theory. MG is supported in part by NSFC Grants No.~11775022 and 11375026. MG also thanks the Perimeter Institute \lq\lq{}Visiting Graduate Fellows\rq\rq{} program. Research at Perimeter Institute is supported by the Government of Canada through the Department of Innovation, Science and Economic Development and by the Province of Ontario through the Ministry of Research, Innovation and Science. The work of NO is supported in part by the project \lq\lq{}Towards
a deeper understanding of black holes with nonrelativistic holography\rq\rq{} of the Independent
Research Fund Denmark (grant number DFF-6108-00340). HY thanks the Theoretical Particle Physics and Cosmology section at the Niels Bohr Institute for support. MG and HY are also financially supported by the China Scholarship Council.

\appendix

\section{Integrals}
\label{app:Integrals}
\subsection{Matched asymptotic expansion method and radial integrals}
\label{app:MAEandRadialIntegral}
The radial integrals appearing in the ``Kerr-MOG lens equations" \eqref{eq:LensEqns} are defined as \cite{gralla2017observational}
\begin{subequations}
\bea
\label{eq:IntergralIr}
I_r&=&M_{\alpha}\int^{r_o}_{r_s}\frac{dr}{\sqrt{\mathcal{R}(r)}},\quad
\tilde{I}_r=2M_{\alpha}\int^{r_s}_{r_{\text{min}}}\frac{dr}{\sqrt{\mathcal{R}(r)}},\\
\label{eq:IntergralJr}
J_r&=&\int^{r_o}_{r_s}\frac{\mathcal{J}_r}{\sqrt{\mathcal{R}(r)}}dr,\quad
\tilde{J}_r=2\int^{r_s}_{r_{\text{min}}}\frac{\mathcal{J}_r}{\sqrt{\mathcal{R}(r)}}dr,
\eea
\bea
\label{eq:MathcalJr}
\mathcal{J}_r&=&\frac{1}{\Delta}\Big[a\big(2M_{\alpha}r-\beta^2
-a\hat{\lambda}\big)-\Omega_s\big(r^4+a^2(r^2+2M_{\alpha}r-\beta^2)\nn\\
&&-a(2M_{\alpha}
r-\beta^2)\hat{\lambda}\big)\Big],
\eea
\end{subequations}
where $M_{\alpha}$ and $\beta$ are defined in \eqref{beta}, $\mathcal{R}(r)$ is defined in \eqref{eq:DefofR},  $r_{\text{min}}$ is the largest (real) root of $\mathcal{R}(r)=0$. These equations are valid when $r_{\text{min}}<r_s$, which is always true for light that can reach infinity. These equations are very similar to the Kerr case \cite{gralla2017observational} but the differences are implied in the specific expressions of $M_{\alpha}$, $\Delta$, $\Omega_s$ and $\mathcal{J}_r$.

We then get the radial integrals in the limit $\epsilon\rightarrow0$ by using the matched asymptotic expansion (MAE) method which was introduced in Ref.~\cite{porfyriadis2017photon,gralla2017observational}.
\subsubsection{First example: $I_r$}
\label{subsec:FirstExample}
We first perform the LHS integral of Eq.~\eqref{eq:IntergralIr}.
We set $M_{\alpha}=1$ in the following. For near-extremal Kerr-MOG black hole, we have (according to \eqref{eq:RecallExpansion})
\begin{subequations}
\label{eq:ParameterExpansion}
\bea
\alpha&=&\frac{1}{a^2}-1+\mathcal{O}(\epsilon^3),\qquad
r_s=1+\epsilon\bar{R},\\
\hat{\lambda}&=&\frac{1+a^2}{a}(1+\epsilon\lambda),\qquad
\hat{q}=\sqrt{4-\frac{1}{a^2}-q^2}
\eea
\end{subequations}

We will use the dimensionless radial coordinate $R=r-1$. We introduce constants $0<p<1$ and $C>0$ to split the integral into two pieces
\be
I_r=\int^{\epsilon^p C}_{\epsilon\bar{R}}\frac{dR}{\sqrt{\mathcal{R}}}+\int^{R_o}_{\epsilon^p C}\frac{dR}{\sqrt{\mathcal{R}}}.
\ee
The scaling of $\epsilon^p$ introduces a separation of scales $\epsilon\ll\epsilon^p\ll1$ as $\epsilon\rightarrow0$, such that the first part of the integral is in the near horizon region $R\sim\epsilon$ and the second  in the far region $R\sim1$.

In the near horizon region, we make the change of variables $x=R/\epsilon$ and expand in $\epsilon$ at fixed $x$. Thus the first part
of the integral is:
\begin{widetext}{2}
\bea
\label{eq:NearRegion}
\int^{\epsilon^p C}_{\epsilon\bar{R}}\frac{dR}{\sqrt{\mathcal{R}}}&=&
\int^{\epsilon^{p-1}C}_{\bar{R}}\frac{dx}{\sqrt{q^2x^2+4(1+a^2)\lambda x+(1+a^2)^2\lambda^2}}+\mathcal{O}(\epsilon)\nn\\
&=&\frac{1}{q}\log\Bigg[\frac{2q^2}{qD_s+q^2
\bar{R}+2(1+a^2)\lambda}+(p-1)\log\epsilon+\log C\Bigg]+\mathcal{O}(\epsilon^p),
\eea
\end{widetext}
where $D_s$ is defined as
\be
\label{eq:Ds}
D_s=\sqrt{q^2\bar{R}^2+4(1+a^2)\lambda\bar{R}+(1+a^2)^2\lambda^2}.
\ee

In the far region, we expand in $\epsilon$ at fixed $R$. The second integral then reads:
\bea
\label{eq:FarRegion}
\int_{\epsilon^p C}^{R_o}\frac{dR}{\sqrt{\mathcal{R}}}&=&\int_{\epsilon^p C}^{R_o}\frac{dR}{R\sqrt{q^2+4R+R^2}}+\mathcal{O}(\epsilon)\nn\\
&=&\frac{1}{q}\log\Bigg[\frac{2q^2R_o}{qD_o+q^2+2R_o}-p\log\epsilon-\log C\Bigg]\nn\\
&&+\mathcal{O}(\epsilon^p),
\eea
where $D_o$ is defined as
\be
\label{eq:Do}
D_o=\sqrt{q^2+4R_o+R_o^2}.
\ee

By adding Eqs.~\eqref{eq:NearRegion} and \eqref{eq:FarRegion}, we get the complete integral:
\bea
I_r&=&\frac{1}{q}
\log\Bigg[\frac{4q^4R_o}{(qD_o+q^2+2R_o)(qD_s+q^2
\bar{R}+2(1+a^2)\lambda)}\Bigg]\nn\\
&&-\frac{1}{q}\log\epsilon+\mathcal{O}(\epsilon).
\eea

\subsubsection{Second example: $\tilde{I}_r$}
Next we perform the RHS integral of Eq.~\eqref{eq:IntergralIr}. This integral is in the near horizon region $R\sim\epsilon$. Introducing $x=R/\epsilon$, we get the larger root of $\mathcal{R}(r)=0$ as:
\be
\label{r min}
 x_{\text{min}}=\frac{1+a^2}{q^2}\big(-2\lambda+|\lambda|\sqrt{4-q^2}\big).
\ee
For $\lambda>0$, the turning point is inside the horizon, so that we should exclude that case. However, since we have the $1/\Delta$ factor in Eq.~\eqref{eq:MathcalJr} (which is meaningless when it goes through the horizon), the integral of $\tilde{J}_r$ does not exist at all in that case which precludes the existence of a valid geodesic trajectory. Therefore, we may still compute the integral regardless whether $\lambda$ is negative or positive.
Then we also set $M_{\alpha}=1$ and get the integral as
\bea
\tilde{I}_r&=&
2\int_{x_{\text{min}}}^{\bar{R}}\frac{dx}{\sqrt{q^2x^2+4(1+a^2)\lambda x+(1+a^2)^2\lambda^2}}+\mathcal{O}(\epsilon)\nn\\
&=&\frac{1}{q}\log\Bigg[\frac{(qD_s+q^2
\bar{R}+2(1+a^2)\lambda)^2}{(1+a^2)^2(4-q^2)\lambda^2}
\Bigg]+\mathcal{O}(\epsilon).
\eea

\subsubsection{List of results}
The remaining radial integrals $J_r$ and $\tilde{J}_r$ (Eq.~\eqref{eq:IntergralJr}) can be obtained using these methods. Note that the integral $\tilde{J}_r$ is only valid when $\lambda<0$ since it does not exist when $\lambda>0$ (see the discussion below \eqref{r min}).
We now list all of the radial integrals appearing in the Kerr-MOG  equations \eqref{eq:LensEqns} as follows:
\begin{widetext}{2}
\begin{subequations}
\bea
I_r&=&-\frac{1}{q}\log\epsilon+\frac{1}{q}
\log\Bigg[\frac{4q^4R_o}{(qD_o+q^2+2R_o)(qD_s+q^2
\bar{R}+2(1+a^2)\lambda)}\Bigg]+\mathcal{O}(\epsilon),\\
\tilde{I}_r&=&\frac{1}{q}\log\Bigg[\frac{(qD_s+q^2
\bar{R}+2(1+a^2)\lambda)^2}{(1+a^2)^2(4-q^2)\lambda^2}
\Bigg]+\mathcal{O}(\epsilon),\\
\label{eq:IntegralJr}
J_r&=&-\frac{a(6+a^2)}{1+a^2}I_r-\frac{a}{1+a^2}(D_o-q)-\frac{4a^2-1}{2a(1+a^2)^2}
\Big(\frac{q\bar{R}}{\lambda}-\frac{D_s}{\lambda}\Big)\nn\\
&&+\frac{2a}{1+a^2}\log\Bigg[\frac{(q+2)^2
\bar{R}}{(D_o+R_o+2)(D_s+2\bar{R}+(1+a^2)\lambda)}\Bigg]+\mathcal{O}(\epsilon),\\
\label{eq:IntegralJrTilde}
\tilde{J}_r&=&-\frac{a(6+a^2)}{1+a^2}\tilde{I}_r-\frac{4a^2-1}{a(1+a^2)^2}
\frac{D_s}{\lambda}+\frac{2a}{1+a^2}\log\Bigg[
\frac{(D_s+2\bar{R}+(1+a^2)\lambda)^2}{(4-q^2)\bar{R}^2}\Bigg]
+\mathcal{O}(\epsilon),
\eea
\end{subequations}
where $D_s$ and $D_o$ are defined in Eqs.~\eqref{eq:Ds} and \eqref{eq:Do}. Next, the variations are
\begin{subequations}
\bea
\frac{\partial I_r}{\partial\lambda}&=&\frac{1}{\lambda}\bigg(\frac{\bar{R}}{D_s}-
\frac{1}{q}\bigg),\qquad
\frac{\partial\tilde{I}_r}{\partial\lambda}=-\frac{2}{\lambda}
\frac{\bar{R}}{D_s},\\
\frac{\partial I_r}{\partial q}&=&-\frac{1}{q}I_r-\frac{1}{q(4-q^2)}\bigg[(8-q^2)\Big(\frac{\bar{R}}{D_s}+\frac{1}{D_o}
-\frac{2}{q}\Big)+\frac{2(1+a^2)\lambda}{D_s}+\frac{2R_o}{D_o}\bigg],\\
\frac{\partial \tilde{I}_r}{\partial q}&=&-\frac{1}{q}\tilde{I}_r+\frac{2}{q(4-q^2)}\Bigg[(8-q^2)\frac{\bar{R}}{D_s}
+2(1+a^2)\frac{\lambda}{D_s}\Bigg],\\
\frac{\partial J_r}{\partial\lambda}&=&-\frac{1}{2aD_s}-\frac{1}{\lambda}\Big(\frac{1+a^2}{a}\frac{
\bar{R}}{D_s}-\frac{a(6+a^2)}{1+a^2}\frac{1}{q}\Big)-\frac{4a^2-1}{2a(1+a^2)^2}\frac
{D_s-q\bar{R}}{\lambda^2},\\
\frac{\partial\tilde{J}_r}{\partial\lambda}&=&\frac{1}{aD_s}+\frac{2(1+a^2)\bar{R}}
{aD_s\lambda}+\frac{(4a^2-1)D_s}{a(1+a^2)^2\lambda^2},\\
\frac{\partial J_r}{\partial q}&=&\frac{a(6+a^2)I_r}{(1+a^2)q}
+\frac{4a^2-1}{2a(1+a^2)^2}\Big(\frac{D_s}{q\lambda}-\frac{\bar{R}}{\lambda}\Big)
+\frac{a}{1+a^2}-\frac{2a(10+a^2)}{(1+a^2)q^2}-
\frac{8a(2+a^2)}{(1+a^2)(4-q^2)q^2}\nn\\
&&+\frac{\big[2a^2(2+a^2)(8-q^2)+4(4-q^2)\big]\bar{R}+(1+a^2)\big[4a^2(2+a^2)+
(4-q^2)\big]\lambda}{2a(1+a^2)(4-q^2)qD_s}\nn\\
&&+\frac{a\big[(8-q^2+2R_o)(6+a^2-q^2)\big]}{(1+a^2)(4-q^2)qD_o},\\
\frac{\partial\tilde{J}_r}{\partial q}&=&\frac{a(6+a^2)\tilde{I}_r}{(1+a^2)q}
-\frac{\big[2a^2(2+a^2)(8-q^2)+4(4-q^2)\big]\bar{R}+(1+a^2)\big[4a^2(2+a^2)+
(4-q^2)\big]\lambda}{a(1+a^2)(4-q^2)qD_s}\nn\\
&&-\frac{(4a^2-1)D_s}{a(1+a^2)^2q\lambda}.
\eea
\end{subequations}
\end{widetext}

\subsection{Angular integrals}
\label{app:AngularIntegrals}
The angular integrals appearing in the ``Kerr-MOG lens equations" \eqref{eq:LensEqns} are defined as \cite{gralla2017observational}
\begin{align}
\label{eq:AngleIntegralMS}
	G^{m,s}_i=
	\begin{cases}
		\hat{G}_i\qquad\qquad&m=0,\\
		mG_i-s\hat{G}_i\qquad&m\ge1,
	\end{cases}
	\qquad
	i\in\cu{t,\theta,\phi},
\end{align}
with
\begin{subequations}
\label{eq:AngleIntegral}
\bea
G_\theta&=&M_{\alpha}\int^{\theta_+}_{\theta_-}
\frac{d\theta}{\sqrt{\Theta(\theta)}},
\\
\hat{G}_\theta&=&M_{\alpha}\int^{\pi/2}_{\theta_o}
\frac{d\theta}{\sqrt{\Theta(\theta)}},\\
G_\phi&=&M_{\alpha}\int^{\theta_+}_{\theta_-}
\frac{\csc^2\theta}{\sqrt{\Theta(\theta)}}d\theta,
\\
\hat{G}_\phi&=&M_{\alpha}\int^{\pi/2}_{\theta_o}
\frac{\csc^2\theta}{\sqrt{\Theta(\theta)}}d\theta,\\
G_t&=&M_{\alpha}\int^{\theta_+}_{\theta_-}
\frac{\cos^2\theta}{\sqrt{\Theta(\theta)}}d\theta,
\\
\hat{G}_t&=&M_{\alpha}\int^{\pi/2}_{\theta_o}
\frac{\cos^2\theta}{\sqrt{\Theta(\theta)}}d\theta,
\eea
\end{subequations}
where $\Theta(\theta)$ is defined in \eqref{eq:DefofTheta} and $\theta_{\pm}$ are roots of it.

We then perform the angular integrals.
Note that we have the same expressions for $\Theta(\theta)$ in both Kerr spacetime and Kerr-MOG spacetime, so that we quote relevant results from Ref.~\cite{gralla2017observational}.
For the near-extremal Kerr-MOG spacetime, we have the expansions \eqref{eq:ParameterExpansion}.
We set $M_{\alpha}=1$ and obtain the integrals as:
\begin{subequations}
\bea
G_{\theta}&=&\frac{2}{a\sqrt{-\mathcal{I}_-}}K\Bigg(
\frac{\mathcal{I}_+}{\mathcal{I}_-}\Bigg)+\mathcal{O}(\epsilon),\\
\hat{G}_{\theta}&=&\frac{1}{a\sqrt{-\mathcal{I}_-}}F\Bigg(\Psi_o\Big|
\frac{\mathcal{I}_+}{\mathcal{I}_-}\Bigg)+\mathcal{O}(\epsilon),\\
G_{\phi}&=&\frac{2}{a\sqrt{-\mathcal{I}_-}}\Pi\Bigg(\mathcal{I}_+\Big|
\frac{\mathcal{I}_+}{\mathcal{I}_-}\Bigg)+\mathcal{O}(\epsilon),\\
\hat{G}_{\phi}&=&\frac{1}{a\sqrt{-\mathcal{I}_-}}\Pi\Bigg(\mathcal{I}_+;\Psi_o\Big|
\frac{\mathcal{I}_+}{\mathcal{I}_-}\Bigg)+\mathcal{O}(\epsilon),\\
G_{t}&=&-\frac{4\mathcal{I}_+}{a\sqrt{-\mathcal{I}_-}}E^{\prime}\Bigg(
\frac{\mathcal{I}_+}{\mathcal{I}_-}\Bigg)+\mathcal{O}(\epsilon),\\
\hat{G}_{t}&=&-\frac{2\mathcal{I}_+}{a\sqrt{-\mathcal{I}_-}}
E^{\prime}\Bigg(\Psi_o\Big|
\frac{\mathcal{I}_+}{\mathcal{I}_-}\Bigg)+\mathcal{O}(\epsilon),
\eea
\end{subequations}
where
\be
\mathcal{I}_{\pm}=\frac{q^2-6}{2a^2}\pm\frac{1}{a^2}
\sqrt{8-3q^2+(4-q^2)a^2+\bigg(\frac{q^2}{2}\bigg)^2},
\ee
and
\be
\Psi_o=\arcsin\sqrt{\frac{\cos^2\theta_o}{\mathcal{I}_+}} \ . 
\ee
Furthermore,  $E^{\prime}(\phi|m)=\partial_m E(\phi|m)$, and $F(\phi|m)$, $E(\phi|m)$, $\Pi(n;\phi|m)$ are the incomplete elliptic integrals of the first, second and third kind, respectively, and $K(m)$, $E(m)$, $\Pi(n|m)$ are the complete elliptic integrals of the first, second and third kind, respectively.


\providecommand{\href}[2]{#2}\begingroup\raggedright\endgroup

\end{document}